\documentclass[journal,editorial,accept,moreauthors,pdftex,12pt,a4paper]{mdpi} 

\usepackage{amsfonts,mathtools}
\usepackage{bm}

\usepackage{color}

\usepackage{soul}
\usepackage{amssymb,amsmath}
\usepackage{graphicx}
\usepackage{subfigure}
\makeatletter
\renewcommand{\@thesubfigure}{\normalsize(\textbf{\alph{subfigure}})}
\makeatother
\usepackage{array, dcolumn, multirow}
\usepackage{hyphenat}
\usepackage{enumitem,caption,booktabs}
\setitemize{parsep=0pt,itemsep=0pt,leftmargin=.4in}
\setenumerate{parsep=0pt,itemsep=0pt,leftmargin=.4in}

\lastpage{\pageref*{LastPage}}
\doinum{10.3390/e16094992}
\pubvolume{16}
\pubyear{2014}
\history{Received: 22 June 2014 / Accepted: 20 August 2014 / Published: 17 September 2014}

\Title{Editorial Comment on the Special Issue of ``Information in Dynamical Systems and Complex Systems''}

\Author{Erik M. Bollt * and Jie Sun }

\address[1]{%
Department of Mathematics, Clarkson University, 8 Clarkson Ave, Potsdam, NY 13699-5815, USA;
E-Mail: sunj@clarkson.edu}

\corres{E-Mail: bolltem@clarkson.edu; \linebreak Tel.: +1-315-268-2395.}

\abstract{
This special issue collects contributions from the participants of the  ``Information in Dynamical Systems and Complex Systems'' workshop, which cover a wide range of important problems and new approaches that lie in the intersection of information theory and dynamical systems. The contributions include theoretical characterization and understanding of the different types of information flow and causality in general stochastic processes, inference and identification of coupling structure and parameters of system dynamics, rigorous coarse-grain modeling of network dynamical systems, and exact statistical testing of fundamental information-theoretic quantities such as the mutual information. The collective efforts reported herein reflect a modern perspective of the intimate connection between dynamical systems and information flow, leading to the promise of better understanding and modeling of natural complex systems and better/optimal design of engineering systems.}


\begin{document}
\vspace{-12pt}

\section{Introduction}

From 18--19 July  2013, a workshop was held, entitled, ``Information in Dynamical Systems and Complex Systems Summer 2013 Workshop'' in Burlington, VT, with Organizers: Erik M. Bollt and \linebreak Jie Sun (Clarkson University).
This special issue of Entropy represents collective thoughts on the workshop themes stated as follows.  Given the modern focus of dynamical systems on coupled oscillators that form complex networks, it is important to move forward and explore these problems from the perspective of information content, flow and causality.  The following general themes were discussed: 

\begin{enumerate}
\item Information flow. In particular transfer entropy has gained a great deal of interest in recent years as future states may be conditioned on past states both with and without access to other stochastic processes as a test through Kullback--Leibler divergence, but recent work suggests that there are possible misinterpretations from the use of transfer entropy for causality inference.
\item Causality, and information signatures of causation. A central question in science is what causes outcomes of interest, and in particular for affecting and controlling outcomes this is even more important. From the perspective of information flow, causation and causal inference becomes particularly poignant. 
\item Symmetries and reversibility, which may be exploited in special circumstances to enlighten understanding of causality,structure, as well as clustering.
\item Scales, hierarchies, lead to understanding of relevance and nested topological partitions when defining a scale from which information symbols are determined. Variational and optimization principles of information in physics, in particular regarding maximum entropy principles that lead to the understanding of underlying laws.
\item Randomness, structure and causality. In some sense randomness may be described as external and unmodelled effects, which we may interpret in the context here as {\it unknown} information.
This leads to: 
hidden states and hidden processes, including such methods as hidden Markov models and more generally Bayesian inference methods. In the context of information content of a dynamical system, such perspective should potentially yield better understanding.
\item Measures and metrics of complexity and information content. The phrase ``{\it complexity}'' is commonly used for a wide variety of systems, behaviors and processes, and yet a commonly agreed description as to what the phrase means is lacking.
\item Physical laws as information filters or algorithms. Since physical laws lead to evolution equations, which from the perspective of this discussion defines evolution from some information state to a new information state, then it can be said that physical laws may be described either as algorithms or information filters that translate states.
\end{enumerate}

To this end the papers in this special issue we now summarize and as we see they point clearly at these key problems.

In the paper by Sarah Marzen and James P. Crutchfield, ``Information Anatomy of Stochastic Equilibria'' (in this issue, Reference~\cite{Marzen2014}) there is a beautiful paper describing many different measures following entropy rate of information content in coupled stochastic processes.  Specifically, considering entropy as a measure of randomness, or said otherwise predictability, the basic question is what must be known about past states of a stochastic process in order to make the future predictable.   Furthermore they discuss the role of memory as stored in casual states.  Their program of description they call ``information anatomy'' wherein several quantities are included: bound information describes information between present and past as a mutual information.  Elusive information quantifies predictable information not captured in the present, in other words how much the past couples to the future, again as an appropriately conditioned mutual information.  This relates to possible hidden states and the specific mode of observation.  Further, ephemeral information links uncertainty in the present to past and present, and bound information links information shared in present and future conditioned on past.  Finally enigmatic information is a joint information of past, present and future.  With these quantities so defined, then there are several immediate equations that link these quantities, and this is so-called the information anatomy.    While these were previously defined by the authors~\cite{James2011}, here it was extended to finite-order non-autonomous Markov processes, and also continuous-valued processes.  Many beautiful and insightful perspectives come in our broader understanding of stochastic processes with these many information theoretic connections between past, present and future in a stochastic process. For example, while the general information anatomy may allow predictability, there are instances described where the present information can actually shield past information from the future.  Other useful results include such as a Markov assumption implies zero elusive entropy, and many other such descriptive results about casual inference.  A good deal of rigorous examples are produced for the Langevin processes.   An exciting observation for future work is stated that an information anatomy perspective leads to a trade-off between information storage and a heat-loss description suggestive of a Maxwell's demon.

In ``Coarse Dynamics for Coarse Modeling: An Example From Population Biology'' by Bush and Mischaikow (in this issue, Reference~\cite{Bush2014}) the goal is to rigorously define a coarse grained graph model of a dynamical system.  Furthermore the model should be robust in the sense that it should remain valid across a subset of parameter variations.  In this sense the model should be structurally stable in some sense, but to do this, a qualitative model is produced and perhaps valid on a subset of the phase space, and furthermore as the title states, it represents a coarser picture of the dynamics.  By coarse the authors mean that on larger scale of the measurements in the phase space, it is valid that there are orbits that behave as the coarse grain dynamics predicts.  In this sense, the modeling reminds of a symbolic dynamical system, and the information theoretic description is implicit, although there are key differences.  While the modeling is admittedly crude, its strength is that it is robust across a parameter range.  The methods of developing a database to rigorously describe the global dynamics across a range of parameters is to develop a Morse decomposition leading to a so-called Hasse diagram~\cite{Arai2009,Bush2012}.  The goal is to identify the recurrent behavior by identifying strongly connected path components.  Eventually this should lead to developing an algebraic topological invariant, namely the Conley index.  For now, these methods can be taken to be an approach of bookkeeping the ``database'' of the robust part of the coarse dynamics.  To demonstrate the computational efficacy of their methods, the authors illustrate with a model of population dynamics on gene regulatory/signal transduction networks.  Further several other interesting dynamical scenarios are demonstrated such as bistability in a population dynamics of fish species.  Perhaps a \linebreak take-away message here could be that a database approach to dynamical systems while coarse makes for a high level summary that could be most useful for certain policy scenarios.

The paper, herein \cite{Pethel2014}, by Pethel and Hahs, ``Exact significance test for mutual information,'' develops a key step that is crucial for any serious data-driven estimator of several of the entropies discussed in this issue and well beyond.  That is the authors develop the first exact null hypothesis significance test for mutual information under the presence of temporal dependencies (Markov structure).  Their null hypothesis is that the mutual information is in fact zero.  This is critical in practice since typically when developing an estimator of mutual information, whether most crudely by binning methods \cite{Schindlera2007} or by more sophisticated kernel methods \cite{Kraskov2004,Tsimpiris2012,Vejmelka2008,Vlachos2010}, often one deals with a great deal of data which computes too many estimated values of mutual information many of which are small.  One must decide if they are simply small but nonzero, or the values are insignificant and therefore should be considered (truncated) to zero. Their approach uses a novel resampling technique leading to surrogates that preserve the Markov structure of the original data (whereas most heuristic methods do not and often lead to fundamental errors). From the surrogate data, the null hypothesis is then extracted. The method is shown to be applicable for general finite-order Markov processes where the order must be first determined from another test (exact significance test for Markov order) developed by the same authors in Reference~\cite{Pethel2014PhysicaD}. This was illustrated through data generated from the logistic maps, by considering optimal order using block entropies compared to that of the next highest order. This work is surely an important step toward the outstanding goal of developing an exact significance test for conditional mutual information, since it is this quantity on which transfer entropy, causation entropy, and the several varieties of entropy part of the information anatomy are based, as found in several of the papers in this issue.

The next paper in this special issue is  ``Infinite Excess Entropy Process with Countable-State Generators'' by Travers and Crutchfield (in this issue, Reference~\cite{Travers2014}).  Therein recalled the definition of   the excess entropy between the complete infinite past and complete infinite future of a stochastic process, this quantity has been used as a classifier as ``finitary'' when finite and otherwise as ``infinitary'' and this is used as a measure of correlation and complexity.   While a popular measure, most studies have focused on finite alphabet systems, most of which have shown to be finitary such as finite-order Markov processes and processes with finite-state hidden Markov models.  It is known that symbolic dynamics at the onset of chaos can be infinitary systems.  Here for the first time, building on a result from ``the Santa Fe process'', examples of countable-state hidden Markov models are constructed that generate finite-alphabet but are in fact infinitary processes.  In so doing not only do they offer first time new examples of this scenario, but also further extend the notions of predictability and therefore entropy connecting past to future.

The paper herein, ``Simultaneous State and Parameter Estimation using Maximum Relative Entropy with Nonhomogenous Differential Equation Constrains," by Giffin and Urniezius, \cite{Giffin2014} takes a new approach to the classic problem of filtering from a dynamical system.  Specifically the problem is given an observed (possibly multi-variate) time-series from a differential equation, and with possibly noise on the observations, but assuming a general form for the differential equation that produced the signal, but with unknown parameter values in the system, can the a ``de-noised" signal and best estimates of the parameters be inferred?  Generally, the problem of on-line estimation is known as filtering, and most popularly the Kalman filters have had great adoption, and to the problem at hand the extended Kalman filter, perhaps with a pre-filtering would come to mind.  But as pointed out here, there are underlying assumptions of memorylessness (Markov) the must be met for the Kalman approach, not to mention difficulties for nonlinear systems and systems with far from equilibrium.  If the question is to estimate states and parameters simultaneously and without a Markov assumption, particularly for a nonlinear system, here a new approach is suggested based on the ``maximum relative entropy" (MrE) formalism.  This allows an assumed model form to together with prior assumptions on state and parameter estimates to be incorporated and then updated on the fly based on observables, and not requiring the Markov assumptions of the traditional methods.  Since the MrE is essentially a constrained optimization method, a Lagrange multiplier approach is standard and shown here to lead to useful and quickly implemented update schemes.  An example with a (low-dimensional) RL circuit with some essential information regarding parameter and model form incorporated into the scheme are shown here to lead to excellent estimates.  Thus here entropic methods lead to a practical and important new direction in an important and widely used method in applied engineering, namely filtering.

Interestingly, as reported in the paper  ``Identifying Chaotic FitzHugh-Nagumo Neurons Using Compressive Sensing'' by by Su, Lai, and Wang (herein, Reference~\cite{Su2014}), another perspective of incorporating prior knowledge/information finds application in the identification of system dynamics. In particular, by assuming complete observability and knowledge of the terms/functions that can potentially arise in the equations under a noise-free setting, the problem of nonlinear identification is converted into a linear one with respect to the basis functions. Then, the resulting linear inverse problem is solved by the well-developed ``$\ell_1$ magic'', {\em i.e.}, minimizing error in terms of the $\ell_1$ norm rather than $\ell_2$. This $\ell_1$ approach was initially developed for the recovery of {\it sparse} signals when the measurement matrix satisfies the {\it uniform uncertainty principle} (sometimes also referred to as the {\it restricted isometry hypothesis})~\cite{Candes2005,Candes2006a,Candes2006b}. Curiously, although it remains unclear whether or not such principle/hypothesis was satisfied for the effective measurement matrix that arises after the nonlinear-to-linear transformation, Reference~\cite{Su2014} still finds the $\ell_1$ solution to be quite satisfactory, as illustrated for the synthetic network of nonlinear FitzHugh-Nagumo (neuron) oscillators with linear coupling~\cite{FitzHugh1961,Nagumo1962}. This encouraging finding raises an interesting question about whether one could in fact extend the established ``uncertainty principles'' to the nonlinear, potentially basis-free scenario.

Clearly as revealed by the information anatomy discussions, the connections between past and future is a most fundamental question in any stochastic process as viewed from an information theory perspective.  Since the development of transfer entropy by Schreiber \cite{Schreiber2000} (also see Reference~\cite{Palus2001}) a great deal of attention has been dealt to questions of how information flows from a stochastic system with (at least) two elements.  In the paper by Butail, Ladu, Spinello, and Porfiri, ``Information Flow in Animal-Robot Interactions'' (in this issue, Reference~\cite{Butail2014}), the questions of how groups of animals or groups of animals and a robot can coordinate their actions is studied with the tools of information theory.  Namely it is asked what information may flow between these agents, as measured by transfer entropy.  This is both an experimental and theoretical study.  An experiment with zebra fish is designed together with a fish analogue robot and the trajectories of each are compared and the possible interactions are measured by transfer entropies.  If transfer entropies are a Kullback--Leibler divergence between fish evolution conditioned on its own past, and conditioned on its own past and that of the external agent - the robot, then this is meant to describe if information from the robot influences the fish.  In fact, there is an information flow found, as measured in bits per time units, and the flow is found to be bi-directional and the significance of the measurements is found to be nontrivial.  This represents a unique and new direction to the broader questions of social interactions but specifically here it may bring to bear a strong method to decide the ``simple'' and fundamental, but hard to answer question of ``who follows whom''.

In the paper ``Identifying Coupling Structure in Complex Systems through the Optimal Causation Entropy Principle'' by Sun, Cafaro, and Bollt herein~\cite{Sun2014Entropy}, the authors discussed the general problem of causality inference in complex systems from time series data based on information-theoretic tools. In particular, the authors reviewed a new theoretical quantity called Causation Entropy introduced in~\cite{Sun2014PhysicaD} and subsequent developments in~\cite{Sun2014arXiv}. Causation Entropy can be used to measure direct and indirect information flow among the components within a complex system by various appropriate conditioning. This was shown rigorously and explicitly in Reference~\cite{Sun2014arXiv}, summarized as the Optimal Causation Entropy Principle which states that the direct causal influences of a given component in a system is the unique minimal set of components that maximizes the (unconditional) Causation Entropy to that component. This allows the development of efficient algorithms to infer cause-and-effect relationships without running into the common issue of having false positives (at least in principle).
In the contributed work, the authors present an application of the Optimal Causation Entropy Principle to infer the direct couplings of the repressilator~\cite{Elowitz2000}, which is a synthetic biological system modeling the dynamics of the concentration of three distinct mRNAs and their corresponding proteins.
The time series data is obtained by applying stochastic perturbations to the system at equilibrium and recording both the perturbations and the system's responses shortly after each such perturbation.
The authors show that by using the algorithms developed based on the Optimal Causation Entropy Principle, the accuracy of coupling inferences becomes higher when more data are available and also increases with higher frequency of sampling and is especially immune to false positives.

Finally, the paper ``Cross-scale interactions and information transfer'' (herein, Reference~\cite{Palus2014}) by Pal\v{u}s focuses on a simply stated but often overlooked and extremely interesting problem: given times series data of a {\it single} system that evolves on multiple time scales, how to measure the direct information flow and detect causal relationships between the dynamics on these different time scales? The recipe proposed by Pal\v{u}s first decomposes the measured signal into a number of modes that correspond to some specific amplitudes and phases (for example, via a wavelet transform). Then, the information flow and therefore potential directed causalities are inferred via the estimation of (conditional) mutual information between the respective phase/amplitude signals. 
The proposed method was benchmarked via a model multifractal process which exhibits information transfer from larger to smaller time scales, and further, applied to a long time series of surface air temperature from various European sites. In the latter case no cross-scale phase--phase interactions were found, while slow climate oscillations were detected to (directly or indirectly) influence the future dynamics of the fast temperature variability from an information-theoretic perspective. It is our sense that the many insightful ideas and discussions presented in Reference~\cite{Palus2014} are likely to motivate future development of a coherence framework to define and identify information flow and causality in general spatial-temporal multi-scale stochastic dynamics.

At the outset of the workshop and as the call to this special issue, several questions were asked, some of which we repeat here, now with a post-script interpretation of hindsight that are cumulatively pulled from both the papers submitted here and also the discussions from Burlington:
\begin{enumerate}
\item Can we develop a general mechanistic description of what renders a real complex system different from a large but perhaps simpler system (particularly from an information theory perspective)?
\begin{itemize}
\item This question could eventually lead to many directions across many scientific fields, and the group agreed that this was a problem worth emphasizing.   
Indeed, the concept is weaved deeply into several of the papers here.  If there is a unifying theme, then the paper  \cite{Bush2014} suggests a skeleton or coarse structure across parameters.  \scalebox{.95}[1.0]{On the other hand the casual structure \cite{Sun2014Entropy}} suggests connections between many elements, but possibly when such structure is hierarchical then this should be considered as complex meaning multi-scaled as opposed to simply a large homogenous system.  Finally, the question of finitary \emph{versus} infinitary examples in \cite{Travers2014} suggests an enumerable difference between different types of complex systems, not to mention the information anatomy \cite{Marzen2014} allows a detailed description of how past, present, and future influence each other, and perhaps when considered also conditioned across spatial scales, then this could be the basis of such an interpretation.
\end{itemize}
\item Can physical laws be defined in an algorithmic and information theoretic manner?
\begin{itemize}
\item Clearly this will be a deeply important connection between physics, physical laws that describe governance of measurable quantities, and the knowledge of states that are governed by these laws.  Since knowledge of states represents information, the specific physical laws that evolve these states is a flow of information, and the information anatomy \cite{Marzen2014}  perhaps gets the bottom of the story of how past, present and future are coupled.  Can the physics be inferred directly from the information theory? This would be a new paradigm to discovery to which we have not yet attained but the question remains extremely important.  Partial steps are made in the sense that causal inferences are discussed in \cite{Sun2014Entropy} and also a parametric approach for assumed model forms is the classic filtering problem as discussed in \cite{Giffin2014}.
\end{itemize}
\item Identify engineering applications, especially those that benefit directly from information theory perspective and methods.
How can this perspective impact design? Can specific control methods be developed that benefit?
\begin{itemize}
\item This is the engineering repeat of the previous question, if engineering means designing a problem with desired properties.  The connection between physics and information theory and an algorithmic formalism between them would suggest that if there is an desired engineering design then a good understanding of the information theory should be beneficial.  Simply from an observational stand point, the study of fish (agents) and robot \cite{Butail2014} has the engineering perspective in that the experimental design allows for studies of how adjusting the robots behavior results in the change of the agents behaviors, and how this influences and influenced by the information flow between them.  This then hints at a possible engineering approach for this general question in the future.
\end{itemize}
\item Can group behaviors and cooperative behaviors such as those of animals and humans be better understood in terms of information theoretic descriptions? What role does hierarchical structures come into play?
\begin{itemize}
\item Clearly the paper by \cite{Butail2014} tackles this question directly by demonstrating that group behavior and cooperation as having an underlying information flow that can be measured and adjusted.  What would remain now is to consider hierarchical aspects of group behaviors, perhaps in coarse grains.
\end{itemize}
\item Can methods designed to identify causal influences be adapted to further adjust and define control strategies for complex systems in biological, social, physical and engineering contexts?
\begin{itemize}
\item As the previous question, this asks how can the information theory perspective be turned around for engineering questions of design.  The casual inference studied in \cite{Sun2014Entropy} gives the formal definitions and methods to define what it means to have control strategies that may causally influence states, but does not yet describe the inverse problems necessary for an engineering control strategy asked here.  How can information flow be a design feature to affect desired and controlled states?  This will be a question surely for future technological innovation.  That said, and with the information flow studies of \cite{Butail2014} and information anatomy discussed in \cite{Travers2014}, all of the elements are here to at least well define the question.  Design of a solution here will likely lead to new technologies not even previously dream of.
\end{itemize}
\item Is there a minimal information description of a dynamical system that will facilitate engineering design? Does approximate description of the formal language suffice for approximate modeling lead to faster and easier design?
\begin{itemize}
\item This question lies underlying essentially all of the papers here in, and central to our discussion at the workshop.  The information in a description may be interpreted as imposing a partition of states on a stochastic system, and once the partition is imposed (see for example Reference~\cite{Bollt2001}) the measure on states may follow from which any information theoretic quantity may be computed.  Therefore the question of minimal information description maps to a question of an infemum of the relevant functional, over all possible partitions, and with respect to all measures on each partition.  This is clearly computationally impossible to approach by brute force but the concept seems likely to lie at the heart of understanding a process.  It is interesting to note that in topological dynamics there exists a variational description of the topological entropy which can be described in terms of the ``maximal entropy measure''~\cite{Bollt2013,Parry1964} which in that setting corresponds to the positive answer this question.  On the other hand, this same question may be interpreted constructively; how much information is needed to build or design a given dynamical system.  This brings the question closer to that of Kolmogorov complexity \cite{Chaitin1966,Kolmogorov1965} which specialized to this question here, is how much information is needed to encode the algorithm that produces the observed data.  That question is in general unanswered. 
\end{itemize}
\end{enumerate}

As a summary statement, it was understood that the validity of the popular approaches of information and entropy measures as a system's probe, description, and summary.
A wide variety of entropic principles have become popular lately in a good deal of recent literature, and therefore at this stage it would be most useful to pause and reconsider which of these quantities imply the various relationships.  A clear voice and a summary of connections between these concepts is crucial for this scientific endeavor to proceed properly and it is hoped that this special issue will help add to this clarity.

\medskip
\vspace{12pt}
\noindent Dr. Erik Bollt
\newline
Dr. Jie Sun
\newline
Guest Editors.

\acknowledgements{Acknowledgments}
We thank  Samuel Stanton from the ARO Complex Dynamics and Systems Program for his ongoing and continuous support.
This work was funded by ARO Grant No. W91 1NF-13-l-0161  and ARO Grant No. 61386-EG.


\conflictofinterests{Conflicts of Interest}
The authors declare no conflict of interest.

\bibliographystyle{mdpi}
\makeatletter
\renewcommand\@biblabel[1]{#1. }
\makeatother


\end{document}